 \documentclass[final,5p,times,twocolumn]{elsarticle}

\newtheorem{lem}{Lemma}
\newproof{pf}{Proof}

\usepackage{amssymb}

\begin{document}

\begin{frontmatter}



\title{A remark on Dickey's stabilizing chain}


\author{Andrei K. Svinin}
\ead{svinin@icc.ru}

\address{Institute for System Dynamics and Control Theory, Siberian Branch of Russian Academy of Sciences, P.O. Box 292, 664033 Irkutsk, Russia}

\begin{abstract}
We observe that Dickey's stabilizing chain can be naturally included into two-dimensional chain of infinitely many copies of
equations of KP hierarchy.
\end{abstract}


\end{frontmatter}

\section{Introduction}

In his article \cite{dickey} Dickey introduced the so-called stabilizing
chain of truncated Kadomtsev-Petviashvili (KP) hierarchies. The latter
can be formulated in the language of formal pseudo-differential dressing
operators (dressing $\Psi$DO's) $W_i=1+\sum_{m=1}^iw_{im}\partial^{-m}$ which are forced to be connected by relations
\begin{equation}
(\partial+u_i)W_i=(\partial+v_{i+1})W_{i+1}
\label{0}
\end{equation}
for $i\geq 0$. One requires that solution of (\ref{0}) such that $w_{ii}$ is not identically equal to zero. Evolution equations of KP hierarchy are given by
\begin{equation}
\partial_s W_i=({\cal Q}_i^s)_{+}W_i-W_i\partial^s,
\label{1}
\end{equation}
for $s\geq 2$, where ${\cal Q}_i\equiv W_i\partial W_i^{-1}$ and $\partial_s\equiv \partial/\partial t_s$. Remember that the subscript $+$ means taking only nonnegative powers of $\partial$ in pseudo-differential operator under consideration. To complete the description of stabilizing chain we need to add evolution equations for ``gluing'' fields $u_i$ and $v_i$:
\begin{equation}
\partial_su_i=
-{\rm res}_{\partial}\left((\partial+u_i)({\cal Q}_i^s)_{+}(\partial+u_i)^{-1}\right),
\label{u}
\end{equation}
\begin{equation}
\partial_sv_i=
-{\rm res}_{\partial}\left((\partial+v_i)({\cal Q}_i^s)_{+}(\partial+v_i)^{-1}\right).
\label{v}
\end{equation}
Remember that ${\rm res}_{\partial}\left(\sum a_m\partial^m\right)\equiv a_{-1}$.
It was shown in \cite{dickey} that equations  (\ref{0}-\ref{v}) are well defined.
Moreover, general solution of stabilizing chain is shown to be given in terms
of Wronskians
\[
W_i=\frac{1}{\tau_{i1}}
\left|
\begin{array}{cccc}
y_{0i}&\cdot\cdot\cdot&y_{i-1,i}&1 \\[0.5cm]
y_{0i}^{\prime}&\cdot\cdot\cdot&y_{i-1,i}^{\prime}&\partial \\[0.5cm]
\vdots&\vdots&\vdots&\vdots \\[0.5cm]
y_{0i}^{(i)}&\cdot\cdot\cdot&y_{i-1,i}^{(i)}&\partial^i
\end{array}
\right|\partial^{-i},
\]
\[
u_i=-\partial\ln\frac{\tau_{i+1,2}}{\tau_{i1}},\;\;\;\;
v_i=-\partial\ln\frac{\tau_{i2}}{\tau_{i1}},
\]
where $\tau_{i1}={\rm Wr}[y_{0i},..., y_{i-1,i}]$ and $\tau_{i2}={\rm Wr}[y_{0,i-1},..., y_{i-1,i-1}]$. By definition, the set
$\{y_{0i},..., y_{i-1,i}\}$ is the basis of the kernel for differential operator $P_i\equiv W_i\partial^i$.
In what follows, we set $y_{i-1,i-1}\equiv y_{i-1,i}^{\prime}$. Functions $y_{kl}$ are forced to be solutions of hierarchy evolution equations
$\partial_s y=\partial^s y$. As is known any analytic solution of this hierarchy can be presented as series over Schur polynomials
\[
y=\sum_{m\geq 0}c_mp_m(x, t_2, t_3,...)
\]
Let us remember that Schur polynomials are defined through the relation
\[
\exp\left(\sum_{s\geq 0}t_sz^s\right)=\sum_{m\geq 0}p_mz^m,\;\;\mbox{where}\;\;t_1=x
\]
and have, in virtue of their definition, following easily verified properties:
\[
p_m(x, 0, 0,...)=x^m/m!\;\;\;\mbox{and}\;\;\;\partial_s p_m=\partial^s p_m=p_{m-s}.
\]

As was shown in \cite{dickey} the sequence $\{\tau_{i1}\}$ has the property
of stabilization with respect to gradation which is defined by the rule: $[t_k]=k$. Namely, if one choose
\[
y_{ki}=(-1)^k\left(p_{i-k-1}+c_1^{(k)}p_{i-k}+c_2^{(k)}p_{i-k+1}+\cdot\cdot\cdot\right),
\]
for $k=0,\ldots, i-1$, then any term of weight $l$ do not depend on $i$ when $i\geq l$. In this case, one says that the sequence $\{\tau_{i1}\}$ has the stable limit. Moreover, with special choice of constants  $c_m^{(k)}$ this stable limit yields expression for Kontsevich integral \cite{itzikson}.

In the next two sections we present our observation that it is quite natural to put equations (\ref{0}-\ref{v}) into two-dimensional chain of KP hierarchies.


\section{Two-dimensional chain of KP hierarchies}

\subsection{Two-dimensional chain of dressing $\Psi$DO's}

Here we construct two-dimensional chain of truncated dressing $\Psi$DO's $\{W_{ij}\}$ related with each other by some suitable relations.

With infinite set of suitable constants $\{c_{kl} : k, l\in\mathbb{Z},\; k\geq 0\}$ we define collection of analytic functions
\begin{equation}
\overline{y}_{kl}=\sum_{m\geq 0}c_{k,l-m}\frac{x^m}{m!}.
\label{2}
\end{equation}
Obviously, by definition,  $\overline{y}_{kl}^{\prime}=\overline{y}_{k,l-1}$.
Let us define
\[
\tau_{ij}\equiv{\rm Wr}[\overline{y}_{0,i+1-j},..., \overline{y}_{i-1,i+1-j}]
\]
and  an infinite set of differential operators
\[
P_{ij}=\partial^i+\sum_{m=1}^iw_{im}^j\partial^{i-m}=\frac{1}{\tau_{ij}}
\left|
\begin{array}{cccc}
\overline{y}_{0,i+1-j}&\cdot\cdot\cdot&\overline{y}_{i-1,i+1-j}&1 \\[0.5cm]
\overline{y}_{0,i+1-j}^{\prime}&\cdot\cdot\cdot&\overline{y}_{i-1,i+1-j}^{\prime}&\partial \\[0.5cm]
\vdots&\vdots&\vdots&\vdots \\[0.5cm]
\overline{y}_{0,i+1-j}^{(i)}&\cdot\cdot\cdot&\overline{y}_{i-1,i+1-j}^{(i)}&\partial^i
\end{array}
\right|
\]
for $i, j\in\mathbb{Z}, i\geq 0$. Require that $w_{ii}^j$ is not identically
equal to zero for any values of $i$ and $j$. This is equivalent to the fact
that $P_{ij}$ has not $y={\rm const}$ as a solution. In what follows, it will
be useful following technical proposition.

\begin{lem}
In virtue of their definition, operators $P_{ij}$ satisfy  equations
\begin{equation}
(\partial+v_{i+1,j})P_{i+1,j}=P_{i+1,j+1}\partial,\;\;\;\;
(\partial+u_{ij})P_{ij}=P_{i+1,j+1}
\label{3}
\end{equation}
with
\begin{equation}
v_{ij}=-\partial\ln\left(\frac{\tau_{i,j+1}}{\tau_{i,j}}\right),\;\;\;\;
u_{ij}=-\partial\ln\left(\frac{\tau_{i+1,j+1}}{\tau_{ij}}\right).
\label{4}
\end{equation}
\end{lem}

\begin{pf}
The first relation in (\ref{3}) follows from the fact that ${\rm ker} P_{i+1,j+1}$ is defined as a linear spanning of derivatives of the
functions which belong to ${\rm ker}\: P_{i+1,j}$. Functions $\overline{y}_{0,i+2-j}^{\prime},\ldots, \overline{y}_{i,i+2-j}^{\prime}$ are linearly independent. Otherwise, $y={\rm const}$ will belong to ${\rm ker} P_{i+1,j}$. Moreover, we have the relation
\[
(\partial+v_{i+1,j})P_{i+1,j}(1)=0
\]
hold. From
\[
P_{i+1,j}(1)=\frac{{\rm Wr}[\overline{y}_{0,i+2-j},\ldots, \overline{y}_{i,i+2-j},1]}{\tau_{i+1,j}}=
(-1)^{i+1}\frac{\tau_{i+1,j+1}}{\tau_{i+1,j}}
\]
we derive expression for $v_{ij}$ in (\ref{4}). The second
relation in (\ref{3}) follows from the fact that all basic functions of
${\mathrm ker} P_{i+1,j+1}$ except for $\overline{y}_{i,i+1-j}$ belong to ${\mathrm ker} P_{ij}$.
In addition, we have
\[
(\partial+u_{ij})P_{ij}(y_{i,i+1-j})=
(\partial+u_{ij})\left(\frac{\tau_{i+1,j+1}}{\tau_{ij}}\right)=0.
\]
The latter gives corresponding expression for $u_{ij}$ in (\ref{4}). Therefore lemma is proved.
\end{pf}

As a consequence of (\ref{3}), we have two equations
\begin{equation}
(\partial+v_{i+1,j})W_{i+1,j}=W_{i+1,j+1}\partial,\;
(\partial+u_{ij})W_{ij}=W_{i+1,j+1}\partial
\label{5}
\end{equation}
for $\Psi$DO's $W_{ij}\equiv P_{ij}\partial^{-i}$. So, we can think of $u_{ij}$ and $v_{ij}$ as ``gluing'' variables which relate $\Psi$DO's of special truncated form on two-dimensional chain. Two formulas in (\ref{5}) define shifts $(i, j)\rightarrow (i, j+1)$ and
$(i, j)\rightarrow (i+1, j+1)$, respectively. As a consequence, of these relations we see that $W_{ij}$ also satisfies the relation
\begin{equation}
(\partial+u_{ij})W_{ij}=(\partial+v_{i+1,j})W_{i+1,j}.
\label{6}
\end{equation}
which manages the shift $(i, j)\rightarrow (i+1,j)$. We see that this equation is nothing else but (\ref{0}). The only difference is that dressing operators $W_i$ in (\ref{6}) are parameterized by additional discrete variable $j$.

\subsection{Two-dimensional chain of KP hierarchies}

We know that if one replaces the basis $\{x^m/m!\}$ by that of Schur polynomials $\{p_m(x, t_2, t_3,...)\}$ in (\ref{2}), that is, 
\[
\overline{y}_{kl}\rightarrow y_{kl}=\sum_{m\geq 0}c_{k,l-m}p_m,
\]
then each $W_{ij}$ automatically will be solution of KP hierarchy (\ref{1}) (see, for example \cite{ohta}), while the sequence of dressing operators along shifts $(i, j)\rightarrow (i+1,j+1)$ and $(i, j)\rightarrow (i, j+1)$, due to (\ref{5}) is nothing else but the semi-infinite 1-Toda lattice (the discrete KP hierarchy) with initial condition $W_{0j}=1$. Then ``gluing'' variables $u_{ij}$ and $v_{ij}$, by their construction, automatically satisfy equations \cite{dickey}
\[
\partial_su_{ij}=-{\rm res}_{\partial}\left((\partial+u_{ij})({\cal Q}_{ij}^s)_{+}(\partial+u_{ij})^{-1}\right),
\]
\[
\partial_sv_{ij}=-{\rm res}_{\partial}\left((\partial+v_{ij})({\cal Q}_{ij}^s)_{+}(\partial+v_{ij})^{-1}\right).
\]

\section{Conclusion}

In this brief note we have shown how Dickey' stabilizing chain (\ref{0}-\ref{v}) can be included into two-dimensional lattice of KP hierarchies. One learns from this presentation, that, the latter in a sense can be viewed as a superposition of two compatible discrete KP hierarchies.




\bibliographystyle{elsarticle-num}



\end{document}